\documentclass{article}
\usepackage{spconf,amsmath,multirow,booktabs}
\usepackage{graphicx}
\usepackage{comment}
\usepackage{xcolor}
\usepackage{graphicx}
\usepackage{subcaption}
\usepackage[normalem]{ulem}
\usepackage{amsfonts}

\ninept

\title{O{\small pen}FEAT: Improving speaker identification by \\ Open-Set Few-Shot  Embedding Adaptation with Transformer}

\name{\parbox{\textwidth}{\centering
Kishan K C$^{1 \dagger}$ \qquad Zhenning Tan$^{2 \dagger}$%
\thanks{$^{\dagger}$Equal contribution. This first author was an intern at Amazon.}
\qquad Long Chen$^{2}$ \qquad Minho Jin$^{2}$ \\
Eunjung Han$^{2}$ \qquad Andreas Stolcke$^{2}$ \qquad Chul Lee$^{2}$}}

\address{$^1$Rochester Institute of Technology, Rochester, NY, USA \\
      $^2$Amazon Alexa AI, Sunnyvale, CA, USA}

\begin{document}
\maketitle

\begin{abstract}
Household speaker identification with few enrollment utterances is an important yet challenging problem, especially when household members share similar voice characteristics and room acoustics. A common embedding space learned from a large number of speakers is not universally applicable for the optimal identification of every speaker in a household. In this work, we first formulate household speaker identification as a few-shot open-set recognition task and then propose a novel embedding adaptation framework to adapt speaker representations from the given universal embedding space to a household-specific embedding space using a set-to-set function, yielding better household speaker identification performance. With our algorithm, \textbf{Open}-set \textbf{F}ew-shot \textbf{E}mbedding \textbf{A}daptation with \textbf{T}ransformer (openFEAT), we observe that the speaker identification equal error rate (IEER) on simulated households with 2 to 7 hard-to-discriminate speakers is reduced by 23\% to 31\% relative. 
\end{abstract}

\begin{figure*}[tb]
  \centering
  \includegraphics[width=0.97\textwidth]{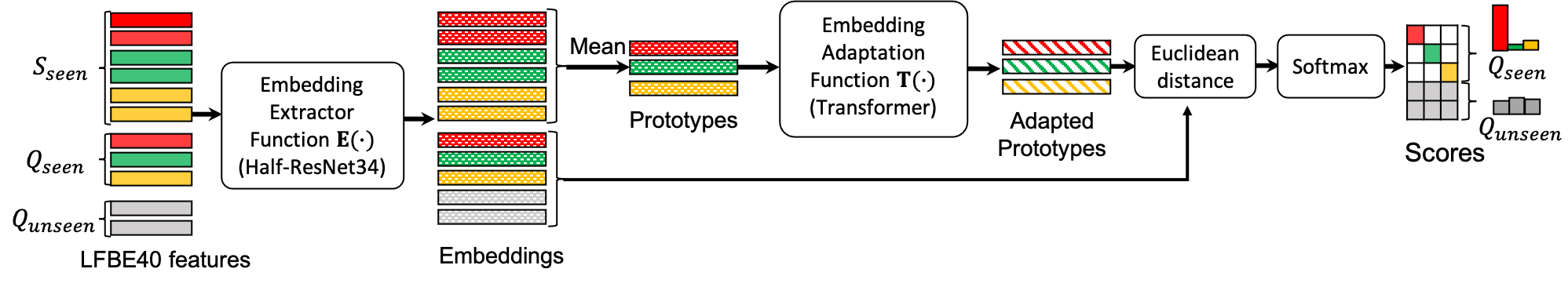}
\vspace{-2mm}
\caption{Architecture of openFEAT} 


\label{fig:emb_adaptation_diagram}
\end{figure*}

\vspace{-2mm}
\section{Introduction}
\vspace{-1mm}

Voice assistants, such as Alexa and Google Home, rely on speaker identification to enable personalized experiences, such as playing one’s favorite music or customizing calendars and alarms. 
On shared devices with multiple enrolled speakers, voice assistants need to distinguish household members based on their speech to identify the user. Since household members often share similar voice characteristics and acoustic conditions, identifying the speaker in this scenario becomes more difficult than for random groups of speakers. On real-world data, we found that the mean cosine similarity between speaker profiles and same-household utterances is about 10\% greater than the similarity with utterances outside the household.  

Speaker identification has two scenarios - closed-set and open-set. In closed-set speaker identification, test utterances come from one of the enrolled speakers \cite{Li2020}. In open-set speaker identification, the test utterance could come from a guest speaker who is not enrolled in the system \cite{fortuna2005open, wilkinghoff2020open}. Open-set speaker identification is closely related to the real-world scenarios where not every utterance comes from an enrolled speaker. 
A typical open-set speaker identification system involves three stages. First, a universal speaker encoder is trained on a large number of speakers to generate speaker or utterance embeddings \cite{chung2020defence}. Second, a function is used to compute the distance between a test utterance embedding and each of the enrolled speaker representations. Finally, the speaker is predicted as the one that is closest to the test utterance based on the similarity measure. The most popular speaker encoders include i-vectors \cite{dehak2010front}, d-vectors \cite{heigold2016end,chung2018voxceleb2,wan2018generalized}, x-vectors \cite{snyder2017deep,snyder2018x}, and many others. The scoring function can be based on probabilistic linear discriminant analysis (PLDA) \cite{ioffe2006probabilistic,prince2007probabilistic}, cosine similarity,
or a neural network \cite{garcia2020magneto,pelecanos2021dr}.

There are two major limitations for the conventional open-set speaker recognition system. First, embeddings learned from the universal speaker encoder are not necessarily optimal for household-level speaker identification, because speech audio from household members often has shared characteristics, such as accents and room acoustics. Moreover, test utterances are compared to speaker profiles in the household \textit{individually}, without considering the similarities shared across speakers in the household. 
Second, the universal speaker encoder is usually trained with classification loss or contrastive loss \cite{chung2020defence}, where query utterances are mapped to one of the classes in the mini batch or episode. The model learns to divide the entire embedding space according to these classes. Unknown classes, which may lie on the decision boundaries, are not considered in this embedding space. The final speaker embeddings are forced into an embedding space that is accurate for seen speakers, which may limit the model's power to generalize well to new speakers.

To address the first limitation, we propose an embedding adaptation by a set-to-set function. It focuses on all speaker profiles \textit{together} in the household and transforms the universal speaker embeddings to household-specific speaker embeddings for better separation. A similar idea of embedding adaptation has been proposed in few-shot learning (FSL) tasks in the computer vision domain \cite{ye2020few}. 

To address the second limitation, we train the embedding adaptation model in a FSL framework for open-set speaker identification task where guest utterances are present during episodic training. In each episode, we have query utterances from both seen classes (with support set utterances) and unseen classes (without support set utterances). For query utterances from unseen classes, we maximize the entropy of the posterior probability distribution with respect to the seen classes to enforce prediction uncertainty. We hypothesize that in this way, the model learns to map speaker embeddings conservatively, considering possible unknown speakers in the entire embedding space. This idea is first proposed in the PEELER framework in the computer vision domain \cite{liu2020few}.

To tackle both problems, we propose a new algorithm, openFEAT, to adapt speaker profile embeddings and to train the model end-to-end in a few-shot open-set recognition framework. To our knowledge, this is the first study to apply embedding adaptation to household speaker profiles to improve speaker recognition and include open-set loss to \textit{train} speaker recognition models to improve generalization of the speaker embedding space. Our contributions are as follow: 1) We demonstrate that few-shot embedding adaptation by transformer (FEAT) and open-set recognition techniques, originally developed in the computer vision domain, can be applied to speaker recognition. 2) We evaluate both techniques and propose a combined solution (openFEAT) that can improve speaker recognition substantially.

\vspace{-2mm}
\section{Proposed Method}
\vspace{-1mm}

\subsection{Few-shot open-set learning (FSOSL)}
In FSL \cite{vinyals2016matching,Sachin2017,snell2017prototypical,finn2017model}, episodes (or tasks) are generated by randomly sampling a subset of the classes and a subset of examples per class. A model is trained on the episodes with classification tasks for which embeddings are optimized, producing robust embeddings (i.e., less overfitting to any particular set of classes). 
To train a speaker embedding extractor, in each episode $D$, we select $N$ speakers with $K$ labeled utterances per speaker, represented as an $N$-way $K$-shot problem. We denote a \textit{support} set as $S_{\text{seen}} = \{x_i^S, y_i^S\}_{i=1}^{NK}$ where $x_i \in \mathcal{X}^S$ represents an input feature for the sample $i$ and $y_i \in \mathcal{Y}^S$ represents its label. We denote a \textit{seen query} set as $Q_{\text{seen}} = \{x_i^Q \in \mathcal{X}^Q, y_i^Q \in \mathcal{Y}^S\}_{i=1}^{NM}$ where $M$ is the number of query utterances for each \textit{seen} speaker.
An embedding extractor function $E(\cdot)$ converts the input features $x_i \in \mathcal{X}^S \cup \mathcal{X}^Q$ to speaker embeddings. With the support embeddings, each class (prototype) is computed by taking an average of the class-wise embeddings: 
$p_c  = \frac{1}{K} \sum_{x \in \mathcal{X}_c^S} E(x)$
where $\mathcal{X}_c^S$ is a set of support instances from class $c$. 
A variety of metric-based FSL methods \cite{schroff2015facenet,snell2017prototypical,wan2018generalized} use softmax classification with a distance function between the query $x$ and the prototypes $P = (p_1, \dots, p_N)$:
\begin{align}\label{softmax}
    f(x, P) = Pr(y=c|x,P) = 
    \frac{e^{-dist(E(x), p_c)}}
    {\sum_{p_i \in P} e^{-dist(E(x), p_i)}}
\end{align}
where $Pr(\cdot)$ is a classification probability, $dist(\cdot, \cdot)$ is a temperature-scaled Euclidean distance \cite{ye2020few}, and $E(\cdot)$ is an embedding extraction function. $E(\cdot)$ is updated by cross-entropy loss as a function of class labels and the class posterior probabilities estimated by the model.

In few-shot open-set learning (FSOSL), in addition to utterances from $N$ seen speakers ($S_\text{seen}$, $Q_\text{seen}$), we randomly select $R$ \textit{unseen} speakers, each with $T$ utterances per episode. We denote  a \textit{unseen query} set as $Q_{\text{unseen}} = \{x_i^U \in \mathcal{X}^U, y_i^U \in \mathcal{Y}^U\}_{i=1}^{RT}$ where $\mathcal{Y}^S \cap \mathcal{Y}^U = \emptyset$. 
For utterances from unseen classes, the model should not assign a large classification probability to any seen classes. To enable this, a loss function should maximize the posterior entropy with respect to the seen classes \cite{liu2020few}. This entropy loss $L_{\text{entropy}}(\cdot)$ for $Q_{\text{unseen}}$ is combined with the cross-entropy loss $L_{\text{CE}}(\cdot, \cdot)$ for $Q_{\text{seen}}$ to form the open-set FSL loss:
\begin{align}
\begin{split}
    \mathcal{L}_{\text{open-set}} =  
    \sum_{(x, y) \in Q_{\text{seen}}} \mathcal{L}_{\text{CE}}(y, f(x, P)) 
    - \beta\sum_{x \in Q_{\text{unseen}}}  \mathcal{L}_{\text{entropy}}(f(x, P))
\end{split}
\end{align}

\vspace{-2mm}
\subsection{Few-shot embedding adaptation with Transformer}
To adapt the prototypes to be more distinguishable in a household-specific space, we train a set-to-set function $T(\cdot)$ which transforms a set of original prototypes $P =(p_1, \dots, p_N)$ to a set of adapted prototypes $P' = (p_1', \dots, p_N')$ where $P' = T(P)$.
We train Transformers \cite{vaswani2017attention,ye2020few} as the set-to-set function:
\begin{align}
\begin{split}
    \tilde{Q} &= W_QP, \tilde{K} = W_KP, \tilde{V} = W_VP, \\ 
    V &= \tilde{V}(\text{softmax}(\frac{\tilde{Q}^T \tilde{K}}{\sqrt{m}})^T), \\
    P' &= \text{LayerNorm}(\text{Dropout}(W_{FC}V) + P) 
\end{split}
\end{align}
where $m$ is the embedding dimension, $W_Q, W_K, W_V \in R^{m \times hm}$ are learnable transform matrices, $h$ is the number of heads, and $W_{FC} \in R^{hm \times m}$ is a  fully connected, trainable weight matrix.
Naturally, the self-attention mechanism without positional encoding allows Transformers to learn contextual relationships between speakers without considering their ordering, which is suitable for the speaker profile adaptation use case.
This algorithm is called Few-Shot Embedding Adaptation with Transformer (FEAT) \cite{ye2020few}. While various set-to-set functions like BiLSTM, DeepSet and Graph Convolution Networks have been studied in \cite{ye2020few}, these do not perform better than FEAT. Thus, in our paper, we only consider FEAT as the algorithm choice for our FSOSL. 

With a transformation function $T(\cdot)$, the classification probability in equation (\ref{softmax}) is now rewritten as
\begin{align}\label{softmax-tr}
    f(x, P') = Pr(y=c|x, P') = 
    \frac{e^{-dist(E(x), p_c')}}
    {\sum_{p_i' \in P'}e^{-dist(E(x), p_i')}}
\end{align}
where $p_c' \in P'$ is the transformed prototype for class $c$.
The embedding adaptation loss is applied to $Q_\text{seen}$ to make sure that the seen query utterances are close to the adapted prototype of the same class and far away from those of other classes:
\begin{align}
\begin{split}
    \mathcal{L}_{\text{query}} =  
    \sum_{(x, y) \in Q_{\text{seen}}} \mathcal{L}_{\text{CE}}(y, f(x, P')) 
\end{split}
\end{align}
Moreover, the embedding adaptation module should  preserve the class-wise similarity after adaptation. To ensure instance embeddings after adaptation are close to their class neighbors and far away from other classes, we apply a contrastive loss as in \cite{ye2020few}. To calculate the contrastive loss, the utterance instances from $Q_\text{seen} \cup S_\text{seen}$ are transformed by the embedding adaptation module instead of prototypes. Then, the new class centers $C$ are calculated with the utterances in $Q_\text{seen} \cup S_\text{seen}$. The adapted instance embeddings are forced to stay close to their class center and far away from other class centers through the cross-entropy loss as:
\begin{align}\label{reg_loss}
    \mathcal{L}_{\text{contrastive}} =
    \sum_{(x, y) \in Q_\text{seen} \cup S_\text{seen}} \mathcal{L}_{\text{CE}}(y, f(x, C))
\end{align}
where $C$ is a set of class centers after the transformation.
Finally, the overall embedding adaptation loss is defined as:
\begin{align}
    \mathcal{L}_{\text{FEAT}} =  \mathcal{L}_{\text{query}} + \alpha \mathcal{L}_{\text{contrastive}} 
\end{align}

During inference, all steps are similar to those for the typical speaker identification system, except that the adapted speaker profiles used for comparison in each household are pre-computed by the embedding adaptation function $T(\cdot)$. Therefore, the embedding adaptation module does not add any run-time computation to the standard speaker identification system.

\vspace{-2mm}
\subsection{openFEAT}
With the previously defined embedding adaptation, speaker profiles are now mapped to a new space. To ensure the newly adapted speaker profiles do not collide with unknown guest speaker embeddings in the new space, we propose to add query utterances from $Q_\text{unseen}$ during embedding adaptation in FSOSL in a similar fashion as that of \cite{liu2020few}. We call our FSOSL algorithm as openFEAT (Figure \ref{fig:emb_adaptation_diagram}). Our openFEAT is based on the intuition that the newly adapted embeddings are only adjusted locally to improve the separability between speaker sets and should not move to a completely different embedding space so as to cause collisions with unseen speakers. The new total loss for openFEAT is: 
\begin{align}
    \mathcal{L}_{\text{openFEAT}} =  
    \mathcal{L}_{\text{query}} + \alpha \mathcal{L}_{\text{contrastive}}  - 
    \beta \sum_{x \in Q_{\text{unseen}}}  \mathcal{L}_{\text{entropy}}(f(x, P'))
\end{align}
where $\alpha$ and $\beta$ are hyper-parameters that control the weight of different losses. Note that the entropy loss for $Q_{\text{unseen}}$ is computed with the adapted prototypes $P'$.

\vspace{-2mm}
\section{Experiments}
\vspace{-1mm}

We used VoxCeleb datasets in our experiments. VoxCeleb2 dev \cite{chung2018voxceleb2} is used to train the embedding function $E(\cdot)$ and the embedding adaptation module $T(\cdot)$. The VoxCeleb1 dataset \cite{Nagrani2017VoxCelebAL} is used for evaluation. There is no overlap between speakers in these two datasets. 

\vspace{-2mm}
\subsection{Household simulation for evaluation}
\label{sec:dataset_prepartion}

To evaluate household level speaker identification, we simulated households with different sizes ($n$). Speakers in the same households may share similar characteristics. To mimic this real-life difficulty, we synthesize households with hard-to-discriminate speakers based on the cosine similarity between the speaker embeddings. Utterance level embeddings are generated using the embedding function $E(\cdot)$ from pretrained models as in Section~\ref{pretraining}, and speaker level embeddings are constructed by averaging up to 100 randomly selected utterances. We then select highly similar speakers, using the 85th percentile among all cosine similarity scores between speaker profiles as a threshold. 

During evaluation, for each speaker, 4 utterances are used as enrollment utterances to create average speaker enrollment profiles, and 10 random utterances are selected as evaluation utterances. In addition, we randomly select $50 \times n$ (where $n$ is the household size) utterances from other speakers outside the household as guest utterances for evaluation. We want the number of guest utterances to be large enough to give reliable performance estimates. With the number chosen here, results for simulated households have small 95\% confidence internals, as shown in Table~\ref{tab:EER_voxcelb_1}. For each enrollment, evaluation or guest utterance, we sample 10 equally spaced 3-second segments and average their embeddings to represent the utterance. For each household size, we run the simulation five times and report the 95\% confidence interval together with the mean of the evaluation metric.

\vspace{-2mm}
\subsection{Pretraining the embedding extractor}
\label{pretraining}
For better initialization for embedding adaptation in FSL, we first pretrain the embedding extractor $E(\cdot)$ on VoxCeleb2 dev data with 800,000 episodes. We chose a half-ResNet34 \cite{heo2020clova}, with half the channels in the residual block as the original ResNet34 \cite{he2016deep}, and with a self-attentive pooling (SAP) layer as $E(\cdot)$. The output embedding dimension is 256. In each episode, we selected $N=400$ speakers, each with $M=1$ support utterance and $K=1$ query utterance, as this was the largest episode that could fit into the GPU memory. We adopted prototypical loss \cite{snell2017prototypical} to learn the embedding extraction function $E(\cdot)$. An Adam optimizer with initial learning rate $10^{-3}$ was used and the learning rate was decayed by a factor of 0.95 every 10 epochs. We used $1 \over 32$ as temperature to scale the similarity measure $dist(\cdot, \cdot)$. This pretrained embedding extractor $E(\cdot)$ was then used to extract embeddings in our household simulation.  Our embedding extractor has 3.2\% equal error rate (EER) on the VoxCeleb1 test set, comparable to what was reported in \cite{chung2020defence}.

\vspace{-2mm}
\subsection{Fine-tune embedding extractor}
In baseline experiments (column ``Baseline'' in 
Table \ref{tab:EER_voxcelb_1}), where only unadapted embeddings are used, we fine-tuned the embedding extractor on the same dataset as used in pretraining using episodes that are close to practical household sizes, for 16000 episodes. For each episode, we selected $N=10$, which is close to, but larger than, typical household sizes so as to ensure sufficient training task difficulty. We set $K=4$ to mimic four enrollment utterances per speaker. Each speaker had $Q=5$ as query utterances. For open-set FSL, we selected an additional $R=5$ speakers, each with $T=5$ query utterances without support utterances. Optimizer, learning rate and its decay schedule are the same as in pretraining.

\vspace{-2mm}
\subsection{Train embedding adaptation}

The embedding adaptation module $T(\cdot)$ was jointly trained with the embedding extractor $E(\cdot)$ using the same episodes as for fine-tuning the embedding extractor. $E(\cdot)$ was initialized with the pretrained weights, while $T(\cdot)$ was randomly initialized.  As the embedding adaptation module we use a Transformer as in \cite{ye2020few}, using a self-attention mechanism with single head and single layer. A dropout rate of 0.5 is applied after a fully connected layer, as in \cite{ye2020few}. The set-to-set function $T(\cdot)$ does not change the embedding dimension. Therefore, the output dimension of $T(\cdot)$ is 256 as well.

\vspace{-2mm}
\subsection{Model evaluation}
During evaluation, we scored utterance similarity using cosines scaled to the interval $[0,1]$. For evaluation utterances in each household, we computed the false acceptance rate (FAR) and false negative identification rate (FNIR). FAR is the fraction of guest utterances that are falsely accepted as enrolled speakers. FNIR is the fraction of enrolled speaker utterances that are misidentified as a different speaker or rejected as a guest speaker. We aggregate false acceptance errors and false negative identification errors and define the identification equal error rate (IEER) as the point when FAR equals FNIR. Thus, our error metric is similar to the EER used in speaker verification, but takes the error types of the speaker identification task into account.

\vspace{-3mm}
\section{Results}

\vspace{-1mm}
\subsection{Embedding adaptation and FSOSL}
\vspace{-2mm}
\vspace{-1mm}
\begin{table}[ht]
\caption{IEER ($\%$) of embedding adaptation models on simulated households (lower is better). $n$ is household size. Numbers in parentheses show relative IEER improvement over baseline. We used $\mathcal{L}_{\text{FEAT}}$ with $\alpha=0.5$ for FEAT, $\mathcal{L}_{\text{open-set}}$ with $\beta = 0.1$ for open-set FSL, and $\mathcal{L}_{\text{openFEAT}}$ with $\alpha = 0.5, \beta = 0.1$ for openFEAT.}
\resizebox{\linewidth}{!}{%
\begin{tabular}{ c | r r r r}
\toprule
\multicolumn{1}{l|}{\textbf{n}} & \multicolumn{1}{c}{\textbf{Baseline}} & \multicolumn{1}{c}{\textbf{FEAT}} & \multicolumn{1}{c}{\textbf{Open-set}} & \multicolumn{1}{c}{\textbf{openFEAT}} \\
\midrule
2  & 6.48$\pm$0.29 & 4.91$\pm$0.15(24.3\%) & 5.16$\pm$0.31(20.4\%) & \textbf{4.49$\pm$0.20(30.7\%)}  \\
3  & 8.65$\pm$0.14 & 6.75$\pm$0.12(22.0\%) & 7.06$\pm$0.21(18.4\%)  & \textbf{6.06$\pm$0.18(30.0\%)}   \\
4  & 10.56$\pm$0.26 & 8.56$\pm$0.12(18.9\%) & 8.73$\pm$0.15(17.4\%)  & \textbf{7.67$\pm$0.21(27.4\%)}  \\
5  & 11.98$\pm$0.18 & 10.01$\pm$0.23(16.5\%) & 10.04$\pm$0.26(16.2\%) & \textbf{9.02$\pm$0.24(24.8\%)}  \\
6  & 13.46$\pm$0.12 & 11.37$\pm$0.12(15.5\%) & 11.23$\pm$0.18(16.5\%)  & \textbf{10.30$\pm$0.23(23.5\%)}  \\
7  & 14.69$\pm$0.37 & 12.45$\pm$0.38(15.3\%) & 12.35$\pm$0.35(16.0\%)  & \textbf{11.35$\pm$0.37(22.8\%)} \\ 
\bottomrule
\end{tabular}}
\label{tab:EER_voxcelb_1}
\end{table}

We evaluated the performance on simulated households and observed that baseline IEER increases with household size, which is expected given that speaker identification becomes harder the more speakers are enrolled. When evaluating the FEAT model, we observe relative improvements of 15.3\% to 24.3\% across all households. For FSOSL, the IEER improvement is 16.0\% to 20.4\% across all households. When combining embedding adaptation with FSOSL, the IEER is reduced further by 22.8\% to 30.7\% relative. 

It is noteworthy that the relative improvement decreases with increasing household size. Since embedding adaptation works by projecting the embeddings to a new space while keeping clusters more separable, we hypothesize that with more speakers in the household, finding such a perfect projection space becomes more difficult; therefore, the improvement levels off with increasing household size. 

We also conducted ablation studies to optimize the hyperparameters $\alpha$ and $\beta$ that control the strength of contrastive learning loss and open-set loss. We find that $\alpha=0.5$ achieves the best embedding adaptation performance for FEAT. Considering both hyperparameters, openFEAT with $\alpha=0.5$ and $\beta=0.1$ achieves the best performance.

\vspace{-2mm}
\subsection{Embedding visualization}
\begin{figure}[tb]
\begin{subfigure}{.49\linewidth}
  \centering
  \includegraphics[width=0.99\linewidth]{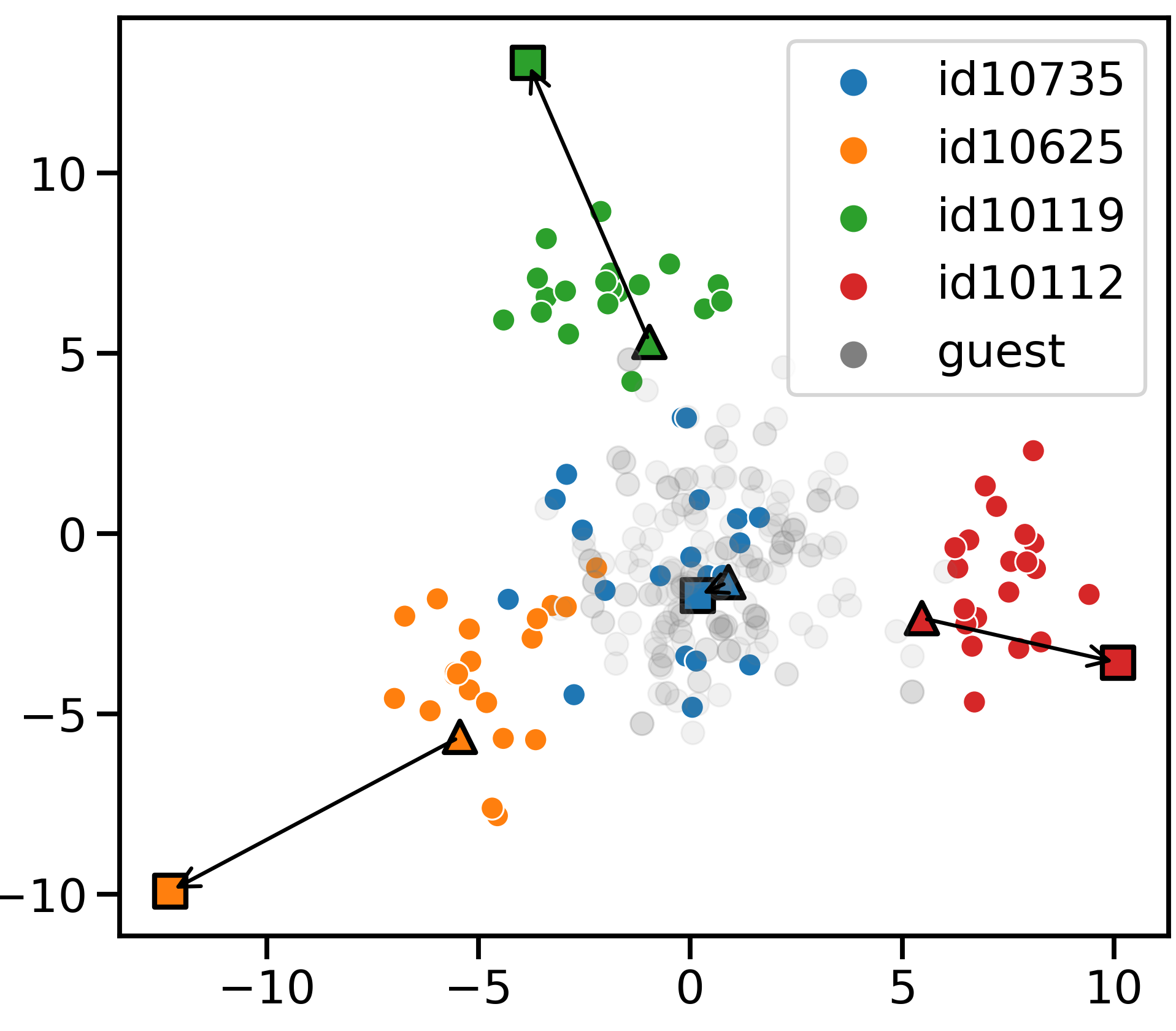}
  \vspace{-5pt}
  \caption{FEAT}
  \label{fig:sub-first}
\end{subfigure}
\begin{subfigure}{.49\linewidth}
  \centering
  \includegraphics[width=0.99\linewidth]{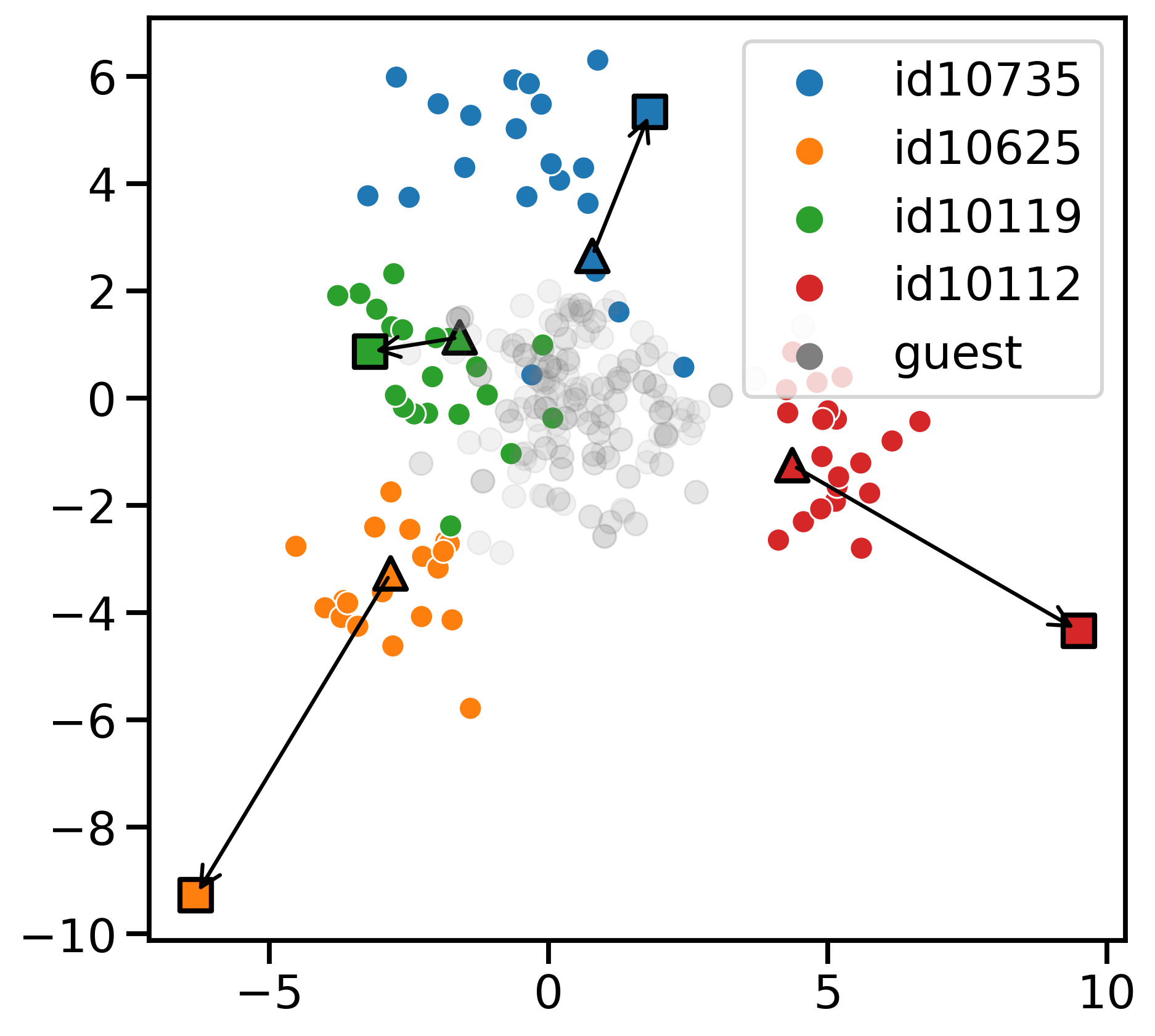} 
  \vspace{-5pt}
  \caption{openFEAT}
  \label{fig:sub-second}
\end{subfigure}
\vspace{-5pt}
\caption{PCA visualization of test utterance embeddings and speaker profiles before and after adaptation in a four-speaker household. Dots represent utterance embeddings. Colored dots are from enrolled speakers, while grey dots are from guest utterances randomly selected from outside the household. Triangles and squares represent the speaker profiles before and after adaptation.}
\label{fig:emb_visualization}
\end{figure}

To qualitatively evaluate the effect of the FEAT embedding adaptation module, we visualized speaker profiles before and after embedding adaptation. We first learn a household-specific PCA projection module using the preadaptation embeddings of the support and query sets for each household. Then, we apply the learned PCA projection to speaker profiles before and after adaptation, along with the evaluation utterance embeddings, both from within and outside the household. As shown in Figure~\ref{fig:emb_visualization}, the adapted speaker profiles are further apart from each other. Interestingly, with openFEAT, we find the speaker profiles can be better separated from guest utterances even along the first two dimensions of the PCA projection.

\vspace{-2mm}
\subsection{Score distribution}
\begin{figure}[tb]
  \centering
  \includegraphics[width=0.9\linewidth]{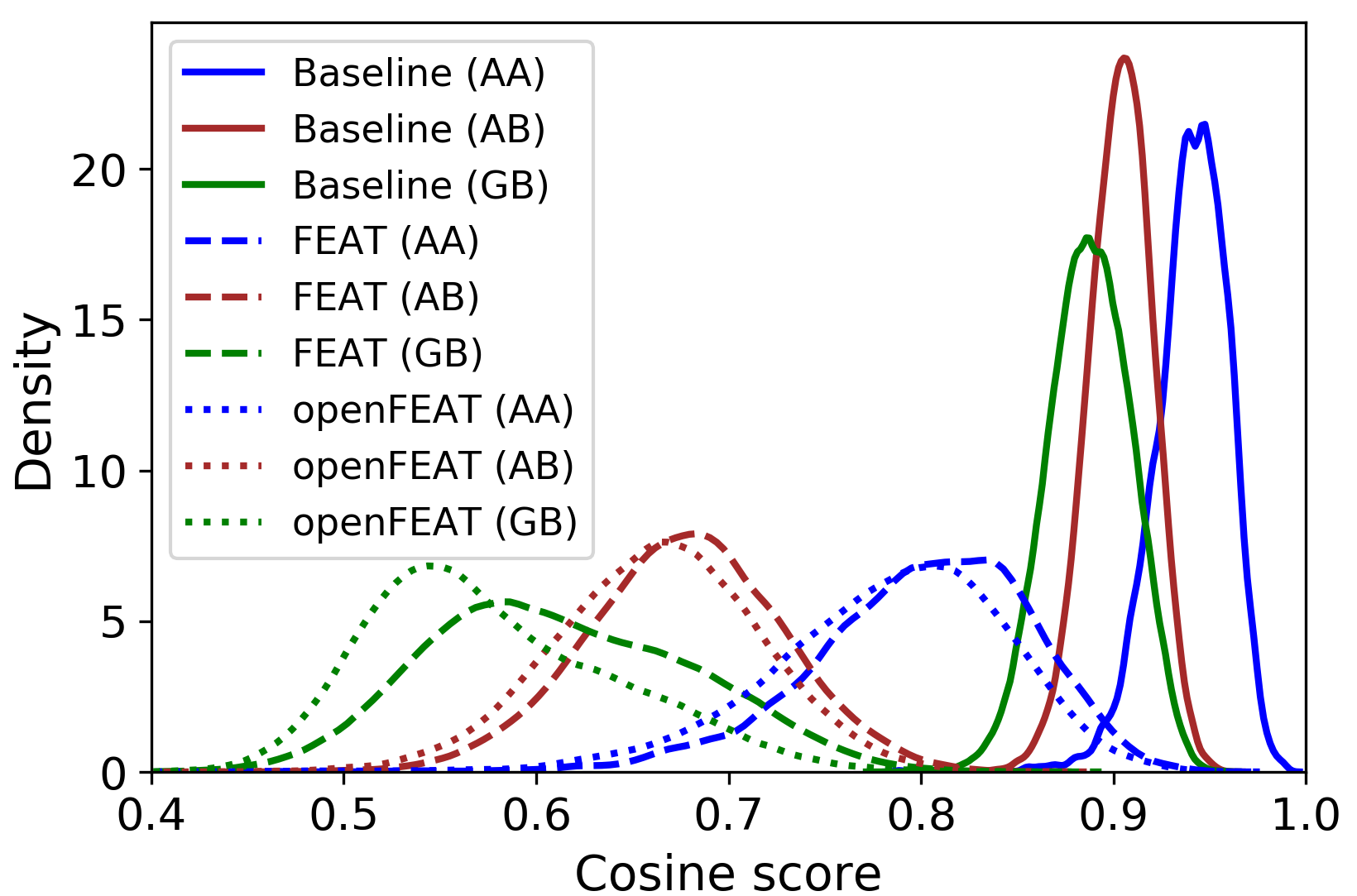}
  \vspace{-5pt}
\caption{Test utterance score distributions for four-speaker households. AA represents the score between an utterance from enrolled speakers and its true speaker. AB represents the score between an utterance from enrolled speakers and its closest other speaker in the household. GB represents the score between a guest utterance and its closest speaker in the household. }
\label{fig:score_distribution}
\end{figure}

We also evaluated changes in the utterance score distribution after speaker profile adaptation. For test utterances from enrolled speaker A, we compute the cosine score between each utterance and its corresponding speaker profile A (AA), as well as to the closest other speaker profile B (AB). For guest utterances (G), randomly selected from outside the household, we only compute the cosine score to the closest speaker profile B (GB) in the household. As shown in Figure \ref{fig:score_distribution}, with embedding adaptation (FEAT), the score distributions have all shifted to a lower score range and the overlap between them is reduced, reflecting lower probability of error. With openFEAT, the guest utterance score distribution is further shifted to a lower range, indicating that the model learns an embedding space for better separation between guest utterances and household speakers.

\vspace{-2mm}
\section{Conclusions}
\vspace{-1mm}

We developed an embedding adaptation method for speaker identification in a household setting, based on Transformer-based set-to-set mapping of profile embeddings. We further incorporated guest utterances into training episodes to reduce speaker embedding confusability with unseen speakers in FSOSL. Our proposed method openFEAT achieves relative IEER reduction of 23\% to 31\% for simulated households of hard-to-discriminate speakers on VoxCeleb1 dataset. Furthermore, we observe that embedding adaptation trained with open-set loss achieves better separation of speaker profiles both in PCA visualizations and score distributions.
\vspace{-2mm}
\section{Acknowledgments}
\vspace{-1mm}

We thank our Speaker Understanding team and Aparna Khare for their valuable feedback, and managers for their inputs and support.

\clearpage
\normalsize
\bibliographystyle{IEEEtran}
\bibliography{refs}

\begin{thebibliography}{10}
\providecommand{\url}[1]{#1}
\csname url@samestyle\endcsname
\providecommand{\newblock}{\relax}
\providecommand{\bibinfo}[2]{#2}
\providecommand{\BIBentrySTDinterwordspacing}{\spaceskip=0pt\relax}
\providecommand{\BIBentryALTinterwordstretchfactor}{4}
\providecommand{\BIBentryALTinterwordspacing}{\spaceskip=\fontdimen2\font plus
\BIBentryALTinterwordstretchfactor\fontdimen3\font minus
  \fontdimen4\font\relax}
\providecommand{\BIBforeignlanguage}[2]{{%
\expandafter\ifx\csname l@#1\endcsname\relax
\typeout{** WARNING: IEEEtran.bst: No hyphenation pattern has been}%
\typeout{** loaded for the language `#1'. Using the pattern for}%
\typeout{** the default language instead.}%
\else
\language=\csname l@#1\endcsname
\fi
#2}}
\providecommand{\BIBdecl}{\relax}
\BIBdecl

\bibitem{Li2020}
R.~Li, J.-Y. Jiang, X.~Wu, C.-C. Hsieh, and A.~Stolcke, ``{Speaker
  Identification for Household Scenarios with Self-Attention and Adversarial
  Training},'' in \emph{Proc.\ Interspeech}, 2020, pp. 2272--2276.

\bibitem{fortuna2005open}
J.~Fortuna, P.~Sivakumaran, A.~Ariyaeeinia, and A.~Malegaonkar, ``Open-set
  speaker identification using adapted {Gaussian} mixture models,'' in
  \emph{Proc.\ 9th European Conference on Speech Communication and Technology},
  2005, pp. 1997--2000.

\bibitem{wilkinghoff2020open}
K.~Wilkinghoff, ``On open-set speaker identification with ivectors,'' in
  \emph{Proc.\ Odyssey Speaker and Language Recognition Workshop}, 2020, pp.
  408--414.

\bibitem{chung2020defence}
J.~S. Chung, J.~Huh, S.~Mun, M.~Lee, H.~S. Heo, S.~Choe, C.~Ham, S.~Jung, B.-J.
  Lee, and I.~Han, ``In defence of metric learning for speaker recognition,''
  in \emph{Proc.\ Interspeech}, 2020, pp. 2977--2981.

\bibitem{dehak2010front}
N.~Dehak, P.~J. Kenny, R.~Dehak, P.~Dumouchel, and P.~Ouellet, ``Front-end
  factor analysis for speaker verification,'' \emph{IEEE Transactions on Audio,
  Speech, and Language Processing}, vol.~19, no.~4, pp. 788--798, 2010.

\bibitem{heigold2016end}
G.~Heigold, I.~Moreno, S.~Bengio, and N.~Shazeer, ``End-to-end text-dependent
  speaker verification,'' in \emph{Proc.\ IEEE ICASSP}, 2016, pp. 5115--5119.

\bibitem{chung2018voxceleb2}
J.~S. Chung, A.~Nagrani, and A.~Zisserman, ``{VoxCeleb2}: Deep speaker
  recognition,'' in \emph{Proc.\ Interspeech}, 2018, pp. 1086--1090.

\bibitem{wan2018generalized}
L.~Wan, Q.~Wang, A.~Papir, and I.~L. Moreno, ``Generalized end-to-end loss for
  speaker verification,'' in \emph{Proc.\ IEEE ICASSP}, 2018, pp. 4879--4883.

\bibitem{snyder2017deep}
D.~Snyder, D.~Garcia-Romero, D.~Povey, and S.~Khudanpur, ``Deep neural network
  embeddings for text-independent speaker verification.'' in \emph{Proc.\
  Interspeech}, 2017, pp. 999--1003.

\bibitem{snyder2018x}
D.~Snyder, D.~Garcia-Romero, G.~Sell, D.~Povey, and S.~Khudanpur,
  ``{X-vectors}: Robust {DNN} embeddings for speaker recognition,'' in
  \emph{Proc.\ IEEE ICASSP}, 2018, pp. 5329--5333.

\bibitem{ioffe2006probabilistic}
S.~Ioffe, ``Probabilistic linear discriminant analysis,'' in \emph{Proc.\
  European Conference on Computer Vision}.\hskip 1em plus 0.5em minus
  0.4em\relax Springer, 2006, pp. 531--542.

\bibitem{prince2007probabilistic}
S.~J. Prince and J.~H. Elder, ``Probabilistic linear discriminant analysis for
  inferences about identity,'' in \emph{Proc.\ IEEE 11th International
  Conference on Computer Vision}, 2007, pp. 1--8.

\bibitem{garcia2020magneto}
D.~Garcia-Romero, G.~Sell, and A.~McCree, ``Magneto: X-vector magnitude
  estimation network plus offset for improved speaker recognition,'' in
  \emph{Proc.\ Odyssey Speaker and Language Recognition Workshop}, 2020, pp.
  1--8.

\bibitem{pelecanos2021dr}
J.~Pelecanos, Q.~Wang, and I.~L. Moreno, ``Dr-vectors: Decision residual
  networks and an improved loss for speaker recognition,'' in \emph{Proc.\
  Interspeech}, 2021, pp. 4603--4607.

\bibitem{ye2020few}
H.-J. Ye, H.~Hu, D.-C. Zhan, and F.~Sha, ``Few-shot learning via embedding
  adaptation with set-to-set functions,'' in \emph{Proc.\ IEEE/CVF Conference
  on Computer Vision and Pattern Recognition}, 2020, pp. 8808--8817.

\bibitem{liu2020few}
B.~Liu, H.~Kang, H.~Li, G.~Hua, and N.~Vasconcelos, ``Few-shot open-set
  recognition using meta-learning,'' in \emph{Proc.\ IEEE/CVF Conference on
  Computer Vision and Pattern Recognition}, 2020, pp. 8798--8807.

\bibitem{vinyals2016matching}
O.~Vinyals, C.~Blundell, T.~Lillicrap, D.~Wierstra \emph{et~al.}, ``Matching
  networks for one shot learning,'' in \emph{Advances in Neural Information
  Processing Systems}, vol.~29, 2016, pp. 3630--3638.

\bibitem{Sachin2017}
S.~Ravi and H.~Larochelle, ``Optimization as a model for few-shot learning,''
  in \emph{Proc.\ International Conference on Learning Representations}, 2017.

\bibitem{snell2017prototypical}
J.~Snell, K.~Swersky, and R.~S. Zemel, ``Prototypical networks for few-shot
  learning,'' \emph{Advances in Neural Information Processing Systems},
  vol.~30, pp. 4077--4087, 2017.

\bibitem{finn2017model}
C.~Finn, P.~Abbeel, and S.~Levine, ``Model-agnostic meta-learning for fast
  adaptation of deep networks,'' in \emph{Proc.\ International Conference on
  Machine Learning}.\hskip 1em plus 0.5em minus 0.4em\relax PMLR, 2017, pp.
  1126--1135.

\bibitem{schroff2015facenet}
F.~Schroff, D.~Kalenichenko, and J.~Philbin, ``Facenet: A unified embedding for
  face recognition and clustering,'' in \emph{Proc.\ IEEE Conference on
  Computer Vision and Pattern Recognition}, 2015, pp. 815--823.

\bibitem{vaswani2017attention}
A.~Vaswani, N.~Shazeer, N.~Parmar, J.~Uszkoreit, L.~Jones, A.~N. Gomez,
  {\L}.~Kaiser, and I.~Polosukhin, ``Attention is all you need,'' in
  \emph{Advances in Neural Information Processing Systems}, 2017, pp.
  5998--6008.

\bibitem{Nagrani2017VoxCelebAL}
A.~Nagrani, J.~S. Chung, and A.~Zisserman, ``{VoxCeleb}: A large-scale speaker
  identification dataset,'' in \emph{Proc.\ Interspeech}, 2017, pp. 2616--2620.

\bibitem{heo2020clova}
H.~S. Heo, B.-J. Lee, J.~Huh, and J.~S. Chung, ``Clova baseline system for the
  {VoxCeleb} speaker recognition challenge 2020,'' \emph{arXiv preprint
  arXiv:2009.14153}, Sep. 2020.

\bibitem{he2016deep}
K.~He, X.~Zhang, S.~Ren, and J.~Sun, ``Deep residual learning for image
  recognition,'' in \emph{Proc.\ IEEE Conference on Computer Vision and Pattern
  Recognition}, 2016, pp. 770--778.

\end{thebibliography}

\end{document}